\documentclass[prl,twocolumn,superscriptaddress]{revtex4-1}
\usepackage{graphicx}
\usepackage{dcolumn}
\usepackage{amsmath,amsthm,amssymb}
\usepackage{subfigure}

\newcommand{\be}{\begin{equation}}
\newcommand{\ee}{\end{equation}}
\newcommand{\beq}{\begin{equation}}
\newcommand{\eeq}{\end{equation}}
\newcommand{\bea}{\begin{eqnarray}}
\newcommand{\eea}{\end{eqnarray}}

\begin{document}
\title{A Kolmogorov-Zakharov Spectrum in $AdS$ Gravitational Collapse}

\author{H. P. de Oliveira}
\affiliation{Universidade do Estado do Rio de Janeiro, Instituto de F\'{\i}sica \\ Departamento de F\'{\i}sica Te\'orica, 20.550-013, Rio de Janeiro, Brazil}
\author{Leopoldo A. Pando Zayas}
\affiliation{Michigan Center for Theoretical
Physics, Randall Laboratory of Physics, The University of
Michigan. Ann Arbor, MI 48109-1120}
\author{E. L. Rodrigues}
\affiliation{Universidade do Estado do Rio de Janeiro, Instituto de F\'{\i}sica \\ Departamento de F\'{\i}sica Te\'orica, 20.550-013, Rio de Janeiro, Brazil}
\date{\today}
\begin{abstract}
We study black hole formation during the gravitational collapse of a massless scalar field in asymptotically $AdS_D$ spacetimes for $D=4,5$.  We conclude that spherically symmetric gravitational collapse in asymptotically $AdS$ spaces is turbulent and  characterized by a Kolmogorov-Zakharov spectrum. Namely, we find that after an initial period of weakly nonlinear evolution, there is a regime where  the power spectrum of the Ricci scalar evolves as $\omega^{-s}$ with the frequency, $\omega$, and $s\approx 1.7\pm0.1$.
\end{abstract}

\maketitle

\noindent \emph{Introduction.}


Due to the AdS/CFT correspondence the question of instability of Anti-de-Sitter ($AdS$) spacetimes sits in the intersection of mathematical and numerical relativity, string theory, field theory and condensed matter physics. The stability of spacetimes is central to the study of general relativity as spacetime  is not just a stage where interactions take place, rather, it is a dynamical participant.  $AdS$ is the maximally symmetric solution to the Einstein equations with a negative cosmological constant and one would expect it to be stable. It has recently been argued that AdS is non-linearly unstable \cite{Bizon:2011gg} to black hole formation. Namely, an arbitrarily small generic perturbation leads to the formation of a black hole. This is completely different from the situation in the other two maximally symmetric solutions to the vacuum Einstein equation -- Minkowski (zero cosmological constant) and de Sitter (positive cosmological constant) -- which are known to be stable.  The feature responsible for this instability is the fact that AdS has a timelike boundary at spatial and null infinity. The mechanism proposed in \cite{Bizon:2011gg}, called weakly turbulent, as responsible for this instability was resonant mode mixing that gives rise to diffusion of energy from low to high frequencies.

The study of non-linear instability of AdS initiated in \cite{Bizon:2011gg} has been followed up in \cite{Jalmuzna:2011qw} for higher dimensions and in \cite{Dias:2011ss} for a different type of configuration in pure $AdS$ but with angular dependence.  Some discussions in the context of the relation to thermalization in the dual field theory have appeared in \cite{Garfinkle:2011tc,Garfinkle:2011hm}.

In this paper we study in more details the turbulent mechanism conjectured in \cite{Bizon:2011gg}.  We implement a suggestion put forward in \cite{deOliveira:2012ac} to apply the arsenal of methods of dynamical systems to the gravitational collapse in asymptotically $AdS$ spacetimes. One of the most celebrated results in the theory of turbulent flows is Kolmogorov's theory of 1941 \cite{K41}, a master piece of dimensional analysis (for modern accounts see the monographs \cite{PopeBook} and \cite{FrischBook}). In \cite{deOliveira:2012ac} we argued that the power spectrum of the Ricci scalar at the origin shows period-doubling, a sign of turbulent evolution; here we establish that, in the appropriate regime, the power spectrum has a Kolmogorov-Zakharov scaling with the frequency, $\omega$, as $\omega^{-s}$, with $s\approx 1.7\pm 0.1$. This spectrum is indicative of wave turbulence, a paradigm describing turbulence among strongly interacting waves \cite{ZLKBook,NazarenkoBook}.

\noindent \emph{Gravitational collapse in asymptotically $D$-dimensional $AdS$ spacetimes}

We consider the dynamics of a massless scalar field, $\varphi$, in $D=d+1$ dimensions minimally coupled to gravity with a negative cosmological constant $\Lambda$:
\begin{equation}
S=\int d^{d+1}x\sqrt{-g}\left(\frac{1}{16\pi G}\left(R-\Lambda\right)-\frac{1}{2}(\partial \varphi)^2\right)
\end{equation}
where $G$ is Newton's constant. We focused on spherically symmetric configurations described by the following metric \cite{Bizon:2011gg},

\small{
\be
ds^2=\sec^2\left(\frac{x}{\ell}\right)\bigg[-A e^{-2\delta} dt^2 + A^{-1}dx^2 +\ell^2\sin^2\left(\frac{x}{\ell}\right)\,d\Omega^2_{d-1}\bigg], \label{eq1}
\ee
}
\noindent where $\ell^2=-d(d-1)/2\Lambda$, $d\Omega^2$ is the metric on the unit $(d-1)$-sphere. The functions $A,\delta$ and the scalar field, $\varphi$, depend on $(t,x)$. Notice that the spatial domain is contained in the interval $0 < x < \pi/2$. The $AdS$ spacetime, which is the maximally symmetric solution to the vacuum Einstein equations with a negative cosmological constant, $\Lambda$, corresponds to $A=1$, $\delta=0$ and $\varphi=0$. In this context it is the analog of the Minkowski spacetime for zero cosmological constant and de Sitter spacetime for positive cosmological constant.

By introducing the auxiliary variables $\Phi=\varphi^\prime$ and $\Pi=A^{-1}\mathrm{e}^\delta \dot{\varphi}$, where the overdots and primes denote derivatives with respect to $t,x$,  respectively, the field equations read:

\bea
\delta^\prime &=& -\frac{8\pi G \ell}{d-1} \cos\left(\frac{x}{\ell}\right)\sin\left(\frac{x}{\ell}\right)(\Pi^2+\Phi^2),\label{eq2} \\
A^\prime &=& -\frac{8\pi G A\ell}{d-1} \cos\left(\frac{x}{\ell}\right)\sin\left(\frac{x}{\ell}\right)(\Pi^2+\Phi^2),\label{eq3} \nonumber \\
+& & \frac{1-A}{\ell\cos\left(\frac{x}{\ell}\right)\sin\left(\frac{x}{\ell}\right)}
\bigg[d-2+2\sin^2\left(\frac{x}{\ell}\right)\bigg] \\
\dot{\Phi}&=&(A\mathrm{e}^{-\delta}\Pi)^\prime, \label{eq4}\\
\dot{\Pi}&=&\frac{1}{\tan^{d-1}\left(\frac{x}{\ell}\right)}
\bigg[\tan^{d-1}\left(\frac{x}{\ell}\right)A\mathrm{e}^{-\delta}\Phi\bigg]^\prime.\label{eq5}
\eea

\noindent The first two equations are constraint equations, the third equation is a consequence of the definition of the auxiliary variables, and the last is the Klein-Gordon equation $g^{\mu\nu}\nabla_\mu(\partial_\nu \varphi)=0$. Hereafter, we assume units where $8\pi G=d-1$.

There is a natural mass function, $m(x,t)$, in $AdS$ spacetimes given by

\be
1 - \frac{2 m}{r^{d-2}} + \frac{r^2}{\ell^2} = g^{\alpha\beta}\partial_\alpha r\,\partial_\beta r, \label{eq6}
\ee

\noindent where the standard spherical coordinate $r$ is related to $x$ as  $r = \ell \tan(x/\ell)$, and one can show that,

\be
m(x,t) = (1-A)\frac{\ell^{d-2}\sin^{d-2}\left(\frac{x}{\ell}\right)}{2 \cos^d\left(\frac{x}{\ell}\right)}.\label{eq7}
\ee

\noindent This expression gives the total mass-energy inside a radius $x$ at the instant $t$. The ADM mass of the system is obtained by evaluating the mass function asymptotically, or $M_{\mathrm{ADM}}=\lim_{x \rightarrow \pi\ell/2} m(x,t)$. As a typical feature of various spacetimes, including $AdS$, the ADM mass is a conserved quantity.

\vspace{0.3cm}
\noindent \emph{Toward a dynamical system via spectral methods}

We integrate numerically the field equations (2)-(5) using the Galerkin-Collocation method \cite{boyd}. The central notion of any spectral method is to approximate a system of partial differential equations by a finite set of ordinary differential equations associated to unknown coefficients or to the grid values of certain quantities. This dynamical system approach proves very useful in describing the complex dynamics of the scalar field collapse in $D$-dimensional $AdS$ spacetimes.

The starting point is to provide an approximation for the metric functions in the form of series of appropriate basis functions:

\bea
\Pi_a&=&\sum_{k=0}^N\,a_k(t) \psi_k(y),\;\;\Phi_a=\sum_{k=0}^{N-1}\,b_k(t) \chi_k(y),\label{eq8}\\
A_a&=&1+\sum_{k=0}^M\,c_k(t) \psi^{(A)}_k(y),\;\;\delta_a=\sum_{k=0}^{M}\,\delta_k(t) \psi^{(\delta)}_k(y).\label{eq9}
\eea

\noindent In the above expressions $M,N$ are the truncation orders; $a_k(t),b_k(t),c_k(t),\delta_k(t)$ are the unknown modes or coefficients that constitute the spectral representation of the metric functions. The basis functions $\psi_k(y),\chi_k(y),\psi^{(A)}_k(y),\psi^{(\delta)}_k(y)$ depend on the spatial coordinate $y=4x/\pi\ell-1$ which varies in the interval $-1 \leq y < 1$. These basis functions are linear combinations of Chebyshev's polynomials which satisfy the condition of regularity of the field equations at the origin ($x=0$ or $y=-1$) and at the spatial infinity ($x=\ell \pi/2$ or $y=1$) \cite{Bizon:2011gg}.






The substitution of the approximations given by Eqs. (\ref{eq8}) and (\ref{eq9}) into the field equations (\ref{eq2})-(\ref{eq5}) results in the residual equations. We impose that the residual equations vanish at certain points known as the collocation or grid points. As a consequence, the field equations are transformed into sets of ordinary differential equations for the modes $(a_k(t),b_j(t))$, along with algebraic equations $c_k=c_k(a_i,b_j,\Pi_l,\Phi_m)$ and $\delta_k=\delta_k(a_i,b_j,\Pi_l,\Phi_m)$, where $\Pi_l,\Phi_m$ are the grid values of the functions $\Pi,\Phi$, respectively.

\begin{figure}[h!]
\begin{center}
\includegraphics[scale=0.4]{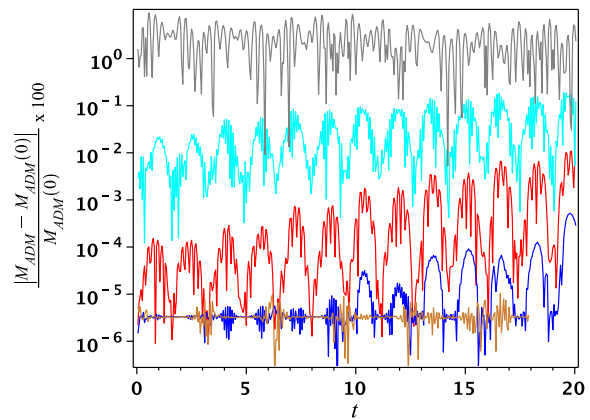}
\vspace{-1cm}
\caption{\label{Error} Relative error of the ADM mass as a function of time.}
\end{center}
\end{figure}

To evolve the spacetime we consider the initial data \cite{Bizon:2011gg}:
$\Phi(0,x)=0$, $\Pi(0,x) = \epsilon_0 \exp\left(-\tan^2x/\sigma^2\right)$, with $\sigma$ and $\epsilon_0$ as free parameters. In the spectral space this initial data is expressed as $b_j(0) =0$ and $\Pi(0,x_j) = \sum_{k=0}^N\,a_k(0) \psi_k(x_j)$, where $x_j$ denotes the collocation points. From this last expression, the initial modes $a_k(0)$ are obtained. With these initial conditions, the dynamical system is numerically evolved and the relevant functions reconstructed afterwards. We will show elsewhere the full details of this procedure and the numerical tests that guarantee the accuracy of the integration. Here, as an example of numerical accuracy, we show in Fig. (\ref{Error}) the behavior of the relative error in the ADM mass $\frac{|M_{ADM}(t)-M_{ADM}(t=0)|}{M_{ADM}(t=0)}$, where $M_{ADM}$ is defined below equation (\ref{eq7}).  The truncation orders are from top to bottom: 10 (gray), 20 (cyan), 30 (red) , 40 (blue) and 50 (gold).Note that the periodic structure in time (with period approximately $\pi$) is the result of the timelike boundary at spatial infinity typical of $AdS$ and has been pointed out in numerical simulations in \cite{Bizon:2011gg,Garfinkle:2011tc,deOliveira:2012ac}.

\noindent \emph{The weakly nonlinear regime}

By decomposing the scalar field in eigenmodes of the $AdS$ Laplacian, the analysis of  \cite{Bizon:2011gg}  gained important insight into the interactions among these modes during the gravitational evolution. It was shown analytically that for generic initial data there is no way to remove all the resonances; some secular terms persist at higher order in perturbation theory, even after accounting for modulation of phase. Inspired by this analytical result, \cite{Bizon:2011gg} also showed numerically that the time scale after which the weakly nonlinear approximation breaks down is of the order $t_{NL}\sim 1/\epsilon_0^2$, where $\epsilon_0$ is the amplitude of the initial profile.

Let us discuss, motivated by the results of \cite{Bizon:2011gg}, a convenient way of viewing the dynamics of a scalar field in $AdS$ spacetimes \cite{deOliveira:2012ac}. In the weak field regime, $\delta,1-A \ll 1$, the Einstein-Klein-Gordon equation in spectral space can be written as,

\small{
\be
\ddot{a}_k(t) - \frac{1}{\tan^2(x(y))}\,[\tan^2(x(y))]^\prime a_k(t) + \mathcal{F}(\delta,1-A,a_k,\dot{a}_k,\ddot{a}_k,..)=0.
\ee}

\noindent The first two terms represent the evolution of the linearized modes (see Wald \cite{wald}) of the scalar field that oscillate in well-defined frequencies $\omega_j^2=(d+2j)^2$, $j=0,1,2,...$. The term  denoted by $\mathcal{F}$ contains the interaction with the gravitational sector. Therefore, we imagine our system as a set of nonlinearly  interacting oscillators, and consequently the periodic motion can be understood as taken place on a torus in the abstract phase space spanned by the modes $a_k(t)$. Further numerical experiments indicated that if the term $\mathcal{F}$ is initially small, its influence on the long term dynamics is to alter the natural frequencies of oscillation producing  quasi-periodic motion. Then, the motion becomes unstable and the torus is eventually destroyed forming a strange attractor, which is a signature of chaos in the  dynamical system. This means that the evolution of the scalar field in AdS is turbulent. This route to turbulence resembles the Ruelle-Takens scenario \cite{ruelle} where the initial amplitude $\epsilon_0$ plays the role of the control parameter. We presented evidence for such picture in \cite{deOliveira:2012ac}.

\noindent \emph{Nonlinear regime: A Kolmogorov-Zakharov spectrum}

For a diffeomorphism invariant theory physical information is contained in coordinate-independent quantities; this is one of the most challenging aspects of applying the standard theory of dynamical systems to gravitational problems. We follow up on one of the suggestions presented in \cite{deOliveira:2012ac}: to study the power spectrum of the scalar curvature. For the backgrounds we are considering the scalar curvature can be written as:
\begin{equation}
R=1/l^2[A\cos(x)^2 (-\Pi^2+\Phi^2)- d(d-1)].
\end{equation}

We have found similar results using the power spectrum of the mass defined in equation \ref{eq7}. Mindful that results about the power spectrum might be affected by coordinate redefinitions but hopeful that the general theorems proving that relativistic chaos is coordinate invariant  \cite{Motter:2003jm} extend to our context, we proceed to a detailed study of properties of the power spectrum of the Ricci scalar.

\begin{figure}[h!]
\begin{center}
\includegraphics[scale=0.33]{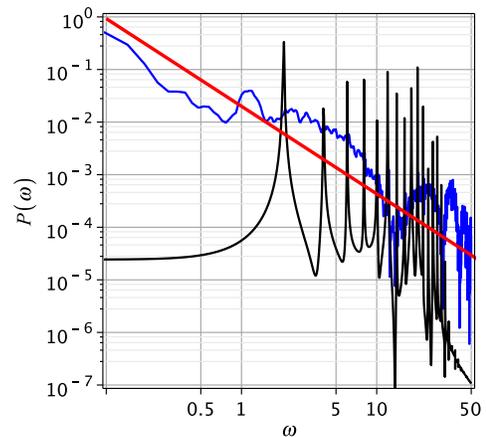}
\caption{\label{Fig.LinTurb} Power spectrum for different amplitudes in gravitational collapse in $AdS_5$. The linear regime ($\epsilon_0=10^{-3}$) with well-defined frequencies (black line); a nonlinear regime ($\epsilon_0=2.5$) that approximately follows a power law and a straight line that is the Kolmogorov $5/3$ scaling (red).}
\end{center}
\end{figure}

Given the notion that for a given amplitude $\epsilon_0$ the weakly nonlinear regime lasts for a time interval proportional to $1/\epsilon_0^2$, by suitably choosing the amplitudes we can reliably separate the regimes of the scalar field evolution during collapse. In Fig. (\ref{Fig.LinTurb}) we present the power spectrum for two systems with amplitudes $10^{-3}$ (red) and $2.5$ (black) respectively. The small amplitude evolution is still in the weakly non-linear regime where the system is well-described by a set of weakly coupled oscillators yielding well-defined peaks at frequencies corresponding to those of the eigenvalues of the Laplace operator in $AdS$. This is precisely what we see in the power spectrum of this signal. Clearly, for higher frequencies we already witness a more complicated picture signalling that the oscillators are not free and, as discussed in \cite{Bizon:2011gg}, and interact by transferring energy to higher frequency modes. In the same figure we have plotted the power spectrum for a higher amplitude simulation in which the nonlinear regimes sets in earlier. The spectrum of the of the nonlinear evolution seems to follow a precise scaling and we have also plotted the -5/3 slope line that corresponds to the Kolmogorov spectrum.

It is well known that Kolmogorov scaling is a property of a
steady-state system, that is, of systems subjected to an injection of
energy at some scale and allowing a sink or energy at some other
scale \cite{K41,PopeBook,FrischBook}. Therefore, the
only way for us to view the appearance of this type of spectrum during collapse is as a regime in the full evolution of the
black hole formation.  Since the time scales of the different regimes is governed by the initial amplitude $\epsilon_0$ we have {\it de facto} created a  steady-state situation by suitably choosing the values of the amplitude.

We are, therefore, not describing a steady-state situation but a phase in the evolution of scalar gravitational collapse in $AdS$. We know that the ultimate fate in the evolution of the system is the formation of a black hole. However, we uncover Kolmogorov scaling by judiciously choosing the amplitude and the time interval from our numerical simulations of gravitational collapse.  We have, at least two important scales in the collapse problem: the nonlinear scale $t_{NL}\sim 1/\epsilon_0^2$ and a new scale that we call the Kolmogorov scale, $t_K$; evolution beyond $t_K$ involves  the violent collapse into a black hole.



\begin{figure}[h!]
\begin{center}
\includegraphics[scale=0.33]{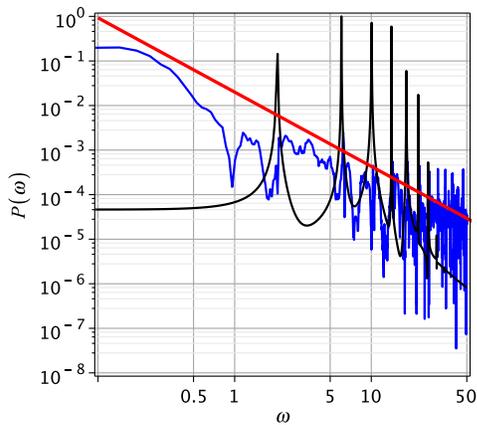}
\caption{\label{Fig.LinTurb_4} Power spectrum for different amplitudes in gravitational collapse in $AdS_4$. The linear regime ($\epsilon_0=10^{-2}$) with well-defined frequencies (black line); a nonlinear regime ($\epsilon_0=2.5$) that approximately follows a power law and a straight line that is the Kolmogorov $5/3$ scaling (red).}
\end{center}
\end{figure}

Turbulence is very difference depending on the dimensionality of the space. Below we provide results for collapse in asymptotically $AdS_4$ spacetimes. However, in the spherically symmetric approximation to gravitational collapse all functions depend on only two coordinates -- time and radius. As can be seen from equation (\ref{eq5}),  different powers of the radius dominate the evolution near the center. We verified explicitly a claim  advanced in  \cite{Jalmuzna:2011qw} that the same turbulent mechanism is at works in spherically symmetric collapse in $AdS_4$ and $AdS_5$. Below we present the power spectrum for gravitational collapse in $AdS_4$ and verify the existence of a power scaling in the appropriate regime see Fig \ref{Fig.LinTurb_4}.

\begin{figure}[h!]
\begin{center}
\includegraphics[scale=0.33]{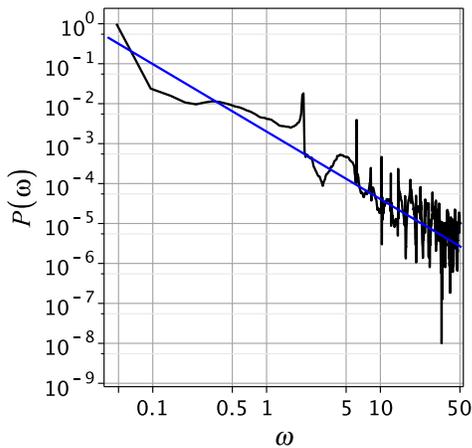}
\caption{\label{InitCond20} Power spectrum for an initial condition $\phi(t=0,x)=\sum\limits_{i=1}^{20} e_i(x)$, built out of 20 eigenmodes of the Laplacian in $AdS$.}
\end{center}
\end{figure}

Another important evidence for the universality of our claim is its independence of the initial scalar field profile. Figure \ref{InitCond20} shows a power spectrum for an initial profile given as the sum of eigenfunctions of $AdS$. Let us briefly explain the reasoning for these initial conditions. The AdS limit of the metric Eq. (\ref{eq1}) corresponds to  $A= 1, \delta = 0$. Expand the scalar field as $\phi(t,x)=\sum\limits_j \phi_j(t) e_j(x)$, where the basis $e_j(x)$ are eigenfunctions of the following operator:
\be
\frac{1}{\tan^2x}\frac{d}{dx}\left(\tan^2x \frac{d}{dx}e_j(x)\right)+(3+2j)^2e_j=0.
\ee
The functions $e_j(x)$ can be orthonormalized as \cite{Bizon:2011gg} :
\bea
e_j(x)&=&d_j \cos^3x\, {}_2F_1(-j,3+j,\frac{3}{2};\sin^2x),  \\
d_j&=&\sqrt{\frac{16(j+1)(j+2)}{\pi}}, \, \int\limits_0^{\pi/2}e_j(x)e_k(x)\tan^2x\, dx = \delta_{jk}. \nonumber
\eea
Upon substitution in the Klein-Gordon equation one obtains
\be
\ddot{\phi_j(t)}+\omega_j^2 \phi_j(t)=0,
\ee
corresponding to free harmonic oscillators with frequencies $\omega_j=3+2j$.  An initial condition of this type is particularly well-suited to the problem and it gives us the ability to study the linearized regime very carefully and understand its transition into the non-linear regime.

We have further performed simulations with a varying number of modes and the results are consistent with those obtained for an initial Gaussian profile. The slope of the power spectrum  corresponds to $s=1.69$,  compatible with those obtained for the Gaussian profile.

\noindent \emph{Conclusions}

We have provided a more complete picture of the gravitational collapse in $AdS$. In particular, we have gone beyond previous descriptions  which  focused on the weakly nonlinear regime characterized by transfer of energy to high frequency modes \cite{Bizon:2011gg}. We have uncovered that after this weakly nonlinear regime a turbulent regimes takes over. Crucially, this turbulent regime is characterized by a Kolmogorov-Zakharov scaling. We have obtained similar results for the power spectrum in $AdS_5$ and $AdS_4$ spacetimes; we have also considered two initial profiles for the scalar field (Gaussian and sum of eigenfunctions). Thus, by considering various dimensions, two types of initial conditions and various parameters of the initial conditions we have found that the range of the power spectrum is $s=1.7\pm 0.1$; we will report on a more detail numerical error analysis elsewhere.

We would like to highlight some of the key open problems that our results motivate.

We have discussed that the steady-state assumption in turbulent flows is achieved in our problem through a separation of scales. Given the connections between gravity and the Navier-Stokes equation \cite{Bredberg:2011jq}, it will be interesting to pursue a description of collapse more along the lines of turbulent fluid dynamics. Namely, we have indications that the phenomena we are witnessing in gravitational collapse is along the lines of wave turbulence
\cite{ZLKBook,NazarenkoBook}. The wave turbulence approach provides an explanation for various aspects that are not typical of full-blown Kolmogorov turbulence such as the similarity of our results in $AdS_5$ and $AdS_4$. More importantly, wave turbulence provides a semi-anlytical approach to our Kolmogorov-Zakharov spectrum. Aspects of wave turbulence in gravitational collapse will be reported elsewhere  \cite{WTCollapse}. Similar applications of wave turbulence have been developed for relativistic system \cite{Micha:2002ey,Micha:2004bv} and provide a tentative explanation of why we find a Kolmogorov-Zakharov (rather than pure Kolmogorov) spectrum in a relativistic system.

It has recently been noted that boson stars might be stable in asymptotically $AdS$ spacetimes because the eigenmodes fail to meet the resonance condition \cite{Dias:2012uy}. It would be interesting to scrutinize this claim under the non-linear perturbation microscope since the very concept of resonance has to be subjected to numerical constraints.

Finally, given that the process of gravitational collapse is dual to thermalization in field theory, it makes sense to speculate that the universality in the turbulent regime means that for strongly coupled field theories the process of thermalization always goes through a similarly turbulent channel.

\noindent \emph{Acknowledgments}
We are thankful to C. A. Terrero-Escalante for collaboration on related topics and to D. Minic and D. Reichmann for important comments and insights. H. P. O. is grateful to  the financial support of Brazilian agencies CNPq and FAPERJ. E. L. R. thanks FAPERJ/CAPES for financial support.  L.A.P.Z. is thankful to  KITP and Aspen Center for Physics for hospitality during various stages of this work.  This research was supported in part by the National Science Foundation under Grant No. NSF PHY11-25915 (KITP),  grant No. 1066293 (Aspen) and by Department
of Energy under grant DE-FG02-95ER40899 to the University of Michigan.


 \end{document}